\documentclass[11pt,superscriptaddress,aps,prd,preprint]{revtex4}
\usepackage[dvips]{graphicx}
\usepackage{amsfonts}
\usepackage{slashed}

\newcommand{\incps}[5]{\includegraphics[#2,#3][#4,#5]{#1}}

\def\tr{\mathrm{tr}}

\def\T{\mathrm{T}}

\def\np{\mathrm{np}}

\begin{document}

\title{Slavnov-Taylor identities for noncommutative QED$_4$}

\author{B. Charneski}
\affiliation{Instituto de F\'{\i}sica, Universidade de S\~ao Paulo\\
Caixa Postal 66318, 05315-970, S\~ao Paulo, SP, Brazil}
\email{bruno,mgomes,ajsilva@fma.if.usp.br}

\author{M. Gomes}
\affiliation{Instituto de F\'{\i}sica, Universidade de S\~ao Paulo\\
Caixa Postal 66318, 05315-970, S\~ao Paulo, SP, Brazil}
\email{bruno,mgomes,ajsilva@fma.if.usp.br}

\author{T. Mariz}
\affiliation{Instituto de F\'\i sica, Universidade Federal de Alagoas, 57072-270, Macei\'o, Alagoas, Brazil}
\email{tmariz@if.ufal.br}

\author{J. R. Nascimento}
\affiliation{Departamento de F\'{\i}sica, Universidade Federal da Para\'{\i}ba\\
Caixa Postal 5008, 58051-970, Jo\~ao Pessoa, Para\'{\i}ba, Brazil}
\email{jroberto@fisica.ufpb.br}

\author{A. J. da Silva}
\affiliation{Instituto de F\'{\i}sica, Universidade de S\~ao Paulo\\
Caixa Postal 66318, 05315-970, S\~ao Paulo, SP, Brazil}
\email{bruno,mgomes,ajsilva@fma.if.usp.br}

\date{\today}

\begin{abstract}
In this work we present an analysis of the one-loop Slavnov-Taylor identities in noncommutative QED$_4$. The vectorial fermion-photon and the triple photon vertex functions were studied, with the conclusion that no anomalies arise.
\end{abstract}

\maketitle

\section{Introduction}

Quantum field theories defined in a noncommutative space have been under intense scrutiny in the last years \cite{Dou,Sza}. The
outcome of these investigations have unveiled various unusual and intriguing  aspects which are consequences of their inherent nonlocality. Among these properties, the most peculiar one is the transmutation of part of the ultraviolet divergences into infrared ones, a property that has been called infrared/ultraviolet mixing \cite{MinRaa}. From a technical viewpoint, the mixing is  due to the separation of the contributions of Feynman diagrams in parts  nonplanar, which are ultraviolet finite but may present an infrared singularity, and planar, which may  have only  ultraviolet divergences. Aside the potentially dangerous character of the  infrared divergences, the mere separation of the  amplitudes in planar and nonplanar parts may obstruct the ultraviolet renormalization of noncommutative theories. 

In the commutative setting, it is well known that Slavnov-Taylor (ST) identities \cite{slavnov} play a fundamental role in the renormalization of non-Abelian gauge theories \cite{slavnov,marciano}. It is therefore  essential to verify to what extension these identities are affected by the noncommutativity of the underlying space. In this work we will present a detailed analysis of the one-loop ST identities in noncommutative QED$_4$. As we will explicitly verify, there are no anomalies and the usual renormalization procedure   is not basically modified.

We would like to point out some relevant studies on the subject. For the pure noncommutative $U(N)$ Yang-Mills model, the compatibility of dimensional renormalization with the ST identities have been verified in \cite{Martin} up to one-loop order. Reference \cite{Jabbari} contains  an explicit on-shell verification of the one-loop ST identity for the trilinear fermion-photon vertex. In the  tree
approximation, the identities have been verified in various scattering processes in \cite{Mariz}. They were also used in \cite{Frenkel} to investigate the dependence of the two point function of the gauge field on the gauge parameter. To prove the absence of radiative corrections to the Chern-Simons coefficient, the axial gauge identities were used and explicitly  verified in a one-loop calculation \cite{Das}.

This work is organized as follows. In section II we introduce our basic notation and the Feynman rules for noncommutative QED$_4$. Section III provides a formal derivation for the ST identities. In particular, using these relations the longitudinal part of photon propagator is fixed and the identities for the vectorial fermion-photon and triple photon vertex functions are presented. In Section IV these identities are subjected to a detailed analysis taking in consideration
the counterterms needed to control the ultraviolet behavior. Section V contains some final comments and a discussion of our results.

\section{Noncommutative QED$_4$}

Classically, the  noncommutative QED$_4$ is described by  the action  
\begin{equation}\label{QED}
S_{INV} = \int d^4x \left[ - \frac14 F_{\mu\nu} \star F^{\mu\nu} + \bar\psi \star (i \slashed{D} - m)\psi \right],
\end{equation}
where $F_{\mu\nu} = \partial_\mu A_\nu - \partial_\nu A_\mu - ie [A_\mu,A_\nu]_\star$, with $[A_\mu,A_\nu]_\star=A_{\mu}\star
A_{\nu}- A_{\nu}\star A_{\mu}$, is the field strength,
$D_\mu \psi = \partial \psi - ie A_\mu \star \psi$  is a gauge covariant derivative and the star (Moyal) product is defined by

\begin{equation}
\phi_{1}(x)\star\phi_{2}(x)\equiv e^{\frac{i}{2}\xi\theta^{\mu
\nu}\frac{\partial\;\;}{\partial x^\mu}\frac{\partial\;\;}{\partial y^\mu}}\phi_{1}(x)\phi_{2}(y)|_{y=x},
\end{equation}
where $\theta_{\mu\nu}$ is a real antisymmetric matrix and $\xi$ is a parameter which sets the strength of the noncommutativity. 

The above action is invariant under the gauge transformations 
\begin{eqnarray}
\delta A_\mu &=&\frac1e  D_{\mu} \Lambda\equiv\frac1e (\partial_\mu \Lambda - ie[A_\mu,\Lambda]_\star), \nonumber\\
\delta \psi &=& i \Lambda \star \psi \quad(\delta \bar\psi = -i \bar\psi \star \Lambda).
\end{eqnarray}

To complete the quantum version of the model,  we need to add to (\ref{QED})  a gauge fixing, $S_{GF}$,  and  the corresponding Faddeev-Popov, $S_{FP}$, actions. For the general class of Lorentz gauges in which we will work 
\begin{equation}
S_{GF} + S_{FP}  = \int d^4 x \left[ - \frac1{2\alpha} (\partial_\mu A^\mu)_\star^2 + \partial_\mu \bar C \star (\partial^\mu C - ie [A^\mu,C]_\star) \right],
\end{equation}
where $\alpha$ is the gauge fixing parameter. As it happens in the commutative gauge theories, the total action $S=S_{INV}+S_{GF}+S_{FP}$ is not invariant under gauge transformations anymore but instead
has a BRST symmetry such that
\begin{eqnarray}
\delta A_\mu &=& -\frac1e (\partial_\mu C - ie[A_\mu,C]_\star) \lambda, \nonumber\\
\delta \psi &=& - i C\lambda \star \psi \quad(\delta \bar\psi = i \bar\psi \star C\lambda),\nonumber\\
\delta C &=& i C \star C\lambda, \nonumber\\
\delta \bar C &=& - \frac{1}{\alpha e} (\partial_\mu A^\mu)\lambda,\label{BRS1}
\end{eqnarray}
where $\lambda$ is a constant Grassmmanian parameter. At a formal level, the invariance of the action under these
transformations imply in relations between the Green functions as it will be shortly verified.  For an explicit
calculation, we will need the Feynman rules for the model which are fixed as follows.
First, the free propagators are the same as in the commutative version of the model, i.e.,
\begin{eqnarray}
\raisebox{-0.4cm}{\incps{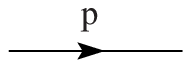}{-1.5cm}{-.5cm}{1.5cm}{.5cm}}
 &=& \frac{i}{\slashed{p}-m}, \\
\raisebox{-0.4cm}{\incps{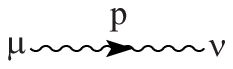}{-1.5cm}{-.5cm}{1.5cm}{.5cm}} &=& -
\frac{i}{p^2} \left[ g^{\mu\nu} - (1-\alpha) \frac{p_\mu p_\nu}{p^2} \right], \;\mathrm{and} \\
\raisebox{-0.4cm}{\incps{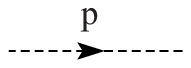}{-1.5cm}{-.5cm}{1.5cm}{.5cm}}
 &=& \frac{i}{p^2},
\end{eqnarray}
for the fermion, photon and ghost field propagators, respectively. Introducing the notation
$p\wedge k \equiv \frac12\xi\theta^{\mu\nu}p_\mu k_\nu$, we determine the
 vertices as being
\begin{eqnarray}
\raisebox{-0.4cm}{\incps{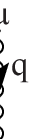}{-1.5cm}{-0.5cm}{1.5cm}{1.5cm}}
 &=& -ie \gamma^\mu e^{i p\wedge k}, \nonumber\\ \\
\raisebox{-0.4cm}{\incps{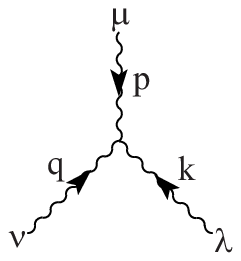}{-1.5cm}{-0.5cm}{1.5cm}{1.5cm}}
 &=& 2e \sin(p\wedge q)\gamma^{\mu\nu\alpha}(p,q,k), \nonumber\\ \\
\raisebox{-0.4cm}{\incps{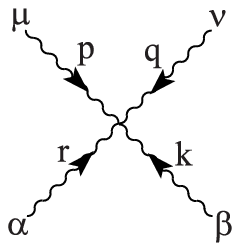}{-1.5cm}{-0.5cm}{1.5cm}{1.5cm}}
 &=& -4ie^2 \left[(g^{\mu\beta}g^{\alpha\nu} - g^{\mu\nu}g^{\alpha\beta})\sin(p \wedge r)\sin(q \wedge k) \right.\nonumber\\
 && + (g^{\mu\nu}g^{\alpha\beta} - g^{\mu\alpha}g^{\nu\beta})\sin(q \wedge p)\sin(r \wedge k) \\
 && + \left.(g^{\mu\alpha}g^{\nu\beta} - g^{\mu\beta}g^{\alpha\nu})\sin(p \wedge k)\sin(r \wedge q)\right], \nonumber\\
\raisebox{-0.4cm}{\incps{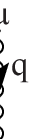}{-1.5cm}{-0.5cm}{1.5cm}{1.5cm}}
 &=& 2e k^\mu \sin(p \wedge k), \nonumber\\
\end{eqnarray}
where $\gamma^{\mu\nu\alpha}(p,q,k)=(p-q)^\alpha g^{\mu\nu}+(q-k)^\mu g^{\nu\alpha}+(k-p)^\nu g^{\alpha\mu}$.

\section{Slavnov-Taylor identities for the generating functionals: formal aspects}

Following the standard procedure adopted in commutative gauge theories,  we start by considering the generating 
functional  for the  Green functions of the basic fields and their BRST variations,
\begin{equation}
Z[J,\eta,\bar\eta,\zeta,\bar\zeta;K,v,\omega,\bar\omega] = \int DA_\mu D\psi D\bar\psi DC D\bar C e^{i(S +S_{source})} \label{Z},
\end{equation}
where $S$ was given in the previous section and
\begin{eqnarray}
S_{source} &=& \int d^4 x \left [J_\mu \star A^\mu + \bar\eta \star \psi + \bar\psi \star \eta + \bar\zeta \star C + \bar C \star \zeta \right. \nonumber\\
&&\left.+ K_\mu \star \frac1e (\partial^\mu C - ie [A^\mu,C]_\star) + iv \star C \star C + i\bar\omega \star C \star \psi + i\bar\psi \star C \star \omega\right].
\end{eqnarray}

The invariance of the functional integral (\ref{Z}) under the field-coordinate transformation (\ref{BRS1}) and the nilpotency of that variations imply  the ST identity
\begin{equation}\label{gfc}
\int d^4x \left(J_\mu \star \frac{\delta W}{\delta K_\mu} - \bar\zeta \star \frac{\delta W}{\delta v} - \frac{1}{\alpha e}\partial_\mu \frac{\delta W}{\delta J_\mu}\star\zeta -\bar\eta \star \frac{\delta W}{\delta\bar\omega} + \frac{\delta W}{\delta\omega} \star \eta\right) = 0,
\end{equation}
where  $W=-i \ln \, Z$ is the generating functional for the connected Green functions. Furthermore, by subjecting  the
functional integral  to an arbitrary variable change  $\delta \bar C$, we may derive that 
\begin{equation}\label{xi}
\zeta = e \partial_\mu \frac{\delta W}{\delta K_\mu}.
\end{equation}

As usual, the generating functional $\Gamma$ of proper (one-particle-irreducible) vertex functions is obtained by a
Legendre transformation
\begin{eqnarray}
W[J,\eta,\bar\eta,\zeta,\bar\zeta;K,v,\omega,\bar\omega] &=& \Gamma[A_{cl},\psi_{cl},\bar\psi_{cl},C_{cl},\bar C_{cl};K,v,\omega,\bar\omega] \nonumber\\
&&+ \int d^4x \left( J_\mu \star A^{\mu}_{cl} + \bar\eta \star \psi_{cl} + \bar\psi_{cl} \star \eta + \bar\zeta \star C_{cl} + \bar C_{cl} \star \zeta \right),
\end{eqnarray}
where we have introduced the classical fields
\begin{equation}
 A^{\mu}_{cl}= \frac{\delta W}{\delta J_\mu},\qquad  \psi_{cl} = \frac{\delta W}{\delta \bar \eta},\qquad \bar \psi_{cl} =- \frac{\delta W}{\delta  \eta},\qquad C_{cl}=\frac{\delta W}{\delta \bar \zeta}, \qquad \bar C_{cl}=-\frac{\delta W}{\delta  \zeta}.
\end{equation}
From these definitions, it follows that 
\begin{equation}
\frac{\delta\Gamma}{\delta A_{cl\,\mu}} = -J^\mu,\;\; \frac{\delta\Gamma}{\delta\psi_{cl}} = \bar\eta,\;\; \frac{\delta\Gamma}{\delta\bar\psi_{cl}} = -\eta,\;\; \frac{\delta\Gamma}{\delta C_{cl}} = \bar\zeta,\;\; \frac{\delta\Gamma}{\delta\bar C_{cl}} = -\zeta.
\end{equation}

In terms of $\Gamma$ the identities (\ref{gfc}) and (\ref{xi}) become
\begin{equation}
\int d^4x \left(\frac{\delta\Gamma}{\delta A^{\mu}_{cl}} \star \frac{\delta\Gamma}{\delta K_\mu} + \frac{\delta\Gamma}{\delta C_{cl}} \star \frac{\delta\Gamma}{\delta v} - \frac{1}{\alpha e}(\partial_\mu A^{\mu}_{cl}) \star \frac{\delta\Gamma}{\delta\bar C_{cl}} + \frac{\delta\Gamma}{\delta\psi_{cl}} \star \frac{\delta\Gamma}{\delta\bar\omega} + \frac{\delta\Gamma}{\delta\omega} \star \frac{\delta\Gamma}{\delta\bar\psi_{cl}}\right) = 0\label{gama}
\end{equation}
and
\begin{equation}
i\frac{\delta\Gamma}{\delta\bar C_{cl}} = - e \partial_\mu \frac{\delta\Gamma}{\delta K_\mu}.
\end{equation}

The identity (\ref{gama}) can be simplified by redefining $\Gamma$:
\begin{equation}
 \Gamma\rightarrow \Gamma- \frac{1}{2\alpha}\int d^4x (\partial_\mu A^{\mu}_{cl})^2
\end{equation}
so that we obtain
\begin{equation}
\int d^4x \left[\frac{\delta\Gamma}{\delta A^{\mu}_{cl}} \star 
\frac{\delta\Gamma}{\delta K_\mu} + \frac{\delta\Gamma}{\delta C_{cl}} 
\star \frac{\delta\Gamma}{\delta v} + \frac{\delta\Gamma}{\delta\psi}
\star \frac{\delta\Gamma}{\delta\bar\omega} +
\frac{\delta\Gamma}{\delta\omega} 
\star \frac{\delta\Gamma}{\delta\bar\psi}\right] = 0
\end{equation}
and
\begin{equation}
i\frac{\delta\Gamma}{\delta\bar C_{cl}} + e \partial_\mu \frac{\delta\Gamma}{\delta K_\mu} = 0.
\end{equation}
Let us now consider some specific applications of the above identities.

\subsection{The photon propagator}

As a first application of the identities derived in the previous section, we will now prove that the longitudinal
part of the photon propagator is not modified by  radiative corrections. To this end, we twice 
differentiate the generating functional of  the connected Green functions (\ref{gfc})  with respect to $\zeta(y)$ and $J^\nu(z)$ and set all sources equal to zero, which gives
\begin{equation}
-\frac1{\alpha e} \partial_y^\mu \frac{\delta^2 W}{\delta J^\nu(z)J^\mu(y)}\left | + \frac{\delta^2 W}{\delta\zeta(y)\delta K^\nu(z)}\right |=0,
\end{equation}
\noindent
where we have introduced the notation ${\cal O}\vert $ to imply that the object ${\cal O}$ at the left of the vertical bar
has to be calculated with all sources equal to zero.
But,  from  Eq. (\ref{xi}) it follows that
\begin{equation}
\partial_z^\nu \frac{\delta^2 W}{\delta\zeta(y)\delta K^\nu(z)}\left | = \frac{1}e \delta(z-y)\right.
\end{equation} 
so that the photon propagator $D_{\mu\nu}(z-y)=-i \frac{\delta^2 W}{\delta J^\nu(z)\delta J^\mu(y)}\left |\right.$ must satisfy
\begin{equation}
\partial_z^\mu \partial_y^\nu D_{\mu\nu}(z-y) =- i\alpha \delta(z-y),
\end{equation}
which  in momentum space becomes
\begin{equation}\label{WTf}
q^\mu q^\nu D_{\mu\nu}(q) =- i\alpha.
\end{equation}
\noindent
Now, compatibility  with this constraint requires the propagator to have the general form
\begin{equation}
 D_{\mu\nu}(q)=\left( g^{\mu\nu} - \frac{q^\mu q^\nu}{q^2} \right) D_\T(q^2) + \frac{\tilde q^\mu\tilde q^\nu}{\tilde q^2} D_\theta(q^2)-\frac{i\alpha}{q^2}\frac{q^\mu q^\nu}{q^2}.\label{1}
\end{equation}
Notice that, because of the  charge conjugation properties \cite{Frenkel,Jabbari1}, terms of the type
\begin{equation}
 \frac{\tilde q^\mu\ q^\nu+\tilde q^\nu\ q^\mu }{\tilde q^2}
\end{equation}
\noindent
are not allowed in the decomposition (\ref{1}). Thus the longitudinal part  of the propagator is the same as in the free approximation. Notice also that
\begin{equation}\label{foton}
q^\mu D_{\mu\nu}(q) = - i\frac{\alpha q_\nu}{q^2},
\end{equation}
which will be useful in the next section when we will analyze the ST identity for the vectorial vertex function.

\subsection{The vectorial vertex function}

The ST identity for the vectorial fermion-photon vertex, the proper part of $\langle 0|T(\psi\bar\psi A_\mu)|0\rangle$, can be derived by turning off all the sources after differentiating the  functional equation (\ref{gfc}) with respect the sources $\eta(y)$, $\bar\eta(x)$, and $\zeta(z)$. The result  is
\begin{equation}
\frac1{\alpha e} \partial_z^\mu \frac{\delta^3 W}{\delta\bar\eta(x)\delta\eta(y)\delta J^\mu(z)}\left | = \frac{\delta^3 W}{\delta\zeta(z)\delta\bar\eta(x)\delta\omega(y)}\right | - \left.\frac{\delta^3 W}{\delta\zeta(z)\delta\eta(y)\delta\bar\omega(x)}\right | ,
\end{equation}
or, equivalently,
\begin{eqnarray}
\frac1{\alpha e} \partial_z^\mu \langle 0|T(\psi(x)\bar\psi(y)A_\mu(z))|0\rangle &=& i \langle 0|T(\bar C(z)\psi(x)\bar\psi(y)\star C(y))|0\rangle \nonumber\\
&&- i \langle 0|T(\bar C(z)\bar\psi(y) C(x)\star\psi(x))|0\rangle,
\end{eqnarray}
i.e.,
\begin{eqnarray}\label{vertex}
\frac1{\alpha e} \partial_z^\mu \langle 0|T(\psi(x)\bar\psi(y)A_\mu(z))|0\rangle &=& i e^{i \partial_x\wedge\partial_{\hat x}} \langle 0|T(\psi(\hat x)\bar\psi(y)C(x)\bar C(z))|0\rangle|_{\hat x=x} \nonumber\\
&& - i e^{i \partial_y \wedge\partial_{\hat y}}\langle 0|T(\psi(x)\bar\psi(y) C(\hat y)\bar C(z))|0\rangle|_{\hat y=y},
\end{eqnarray}
where $\partial_x\wedge\partial_{\hat x}=\frac12\xi\theta^{\mu\nu}\frac{\partial\;\;}{\partial x^\mu}\frac{\partial\;\;}{\partial \hat x^\nu}$. Notice that as consequence of this identity  $$\delta \langle 0|T(\psi(x)\bar\psi(y)\bar C(z))|0\rangle = 0.$$ We may translate the above equations into identities for the proper, one-particle irreducible, vertex functions. These functions are given  by
\begin{eqnarray}
&& \langle 0|T(\psi(x)\bar\psi(y)A_\mu(z))|0\rangle \nonumber\\
&&\quad\quad\quad\quad = - \int d^4x'd^4y'd^4z' S_F(x-x')\Gamma^\nu(x',y',z')S_F(y'-y)D_{\mu\nu}(z-z'), \\
&& i e^{i \partial_x \wedge\partial_{\hat x}} \langle 0|T(\psi(\hat x)\bar\psi(y)C(x)\bar C(z))|0\rangle|_{\hat x=x} \nonumber\\
&&\quad\quad\quad\quad = \int d^4y'd^4z' H_1(x,y',z')S_F(y'-y)\Delta(z'-z), \\
&&i e^{i \partial_y \wedge\partial_{\hat y}} \langle 0|T(\psi(x)\bar\psi(y) C(\hat y)\bar C(z))|0\rangle|_{\hat y=y} \nonumber\\
&&\quad\quad\quad\quad = \int d^4x'd^4z' S_F(x-x')H_2(x',y,z')\Delta(z'-z),
\end{eqnarray}
where  $S_F(x-x')$ and $\Delta(z'-z)$ are the fermion and ghost fields propagators, respectively, $\Gamma^\nu(x',y',z')$ is the vectorial proper vertex,
\begin{equation}
\Gamma^\nu(x',y',z') = \frac{\delta^3 \Gamma}{\delta\bar\psi_{cl}(x')\delta\psi_{cl}(y')\delta A_{cl\nu}(z')},
\end{equation}
\begin{equation}
H_1(x,y',z') = i \int d^4u\,d^4v\, e^{i \partial_x \wedge\partial_{\hat x}} S_F(x-u)\Delta(\hat x-v)\Gamma(u,y',v,z')\vert_{\hat x=x}
\end{equation}  
and
\begin{equation}
H_2(x',y,z') = i \int d^4u\,d^4v\, \Gamma(x',u,v,z') e^{i \partial_y \wedge\partial_{\hat y}} S_F(u-y)\Delta(\hat y-v)\vert_{\hat y=y},
\end{equation}
in which 
\begin{equation}
\Gamma(u,y',v,z') =\left. \frac{\delta^4 \Gamma}{\delta\bar\psi_{cl}(u)\delta\psi_{cl}(y')\delta\bar C_{cl}(v)\delta C_{cl}(z')}\right |
\end{equation}
is the fermion-ghost four-vertex.

In momentum space, Eq. (\ref{vertex}) reads
\begin{equation}\label{WTvertex}
\Gamma^\nu(k,p,q)q^\mu D_{\mu\nu}(-q) = -i \alpha e \left[S_{F}^{-1}(k)H_1(k,p,q)\Delta(q) - H_2(k,p,q)S_{F}^{-1}(p)\Delta(q) \right],
\end{equation}
where
\begin{equation}
H_1(k,p,q) = i \int \frac{d^4k'}{(2\pi)^4} e^{ik'\wedge k}S_F(k')\Delta(k-k')\Gamma(k',p,k-k',q)
\end{equation}
and
\begin{equation}
H_2(k,p,q) = i \int \frac{d^4p'}{(2\pi)^4} e^{-i p'\wedge p}\Gamma(k,p',p'-p,q)S_F(p')\Delta(p'-p),
\end{equation}
with $k=p+q$.

Similarly, we may determine  $\Delta(q)$ from the Dyson-Schwinger equation,
\begin{equation}
\Delta(q) = \Delta_{(0)}(q) - \Delta_{(0)}(q) \Sigma_{C}(q) \Delta(q),
\end{equation}
where $\Delta_{(0)}(q)=\frac{i}{q^2}$ and $\Sigma_{C}(q)$ denotes the proper self-energy operator of the ghost field. Therefore, it is easy to verify that
\begin{equation}\label{ghost}
\Delta(q) = \frac{i}{q^2[1+b(q^2)]},
\end{equation}
in which the self-energy has been expressed as $i\Sigma_{C}(q)=q^2 b(q^2)$.

Thus, with the expressions (\ref{foton}) and (\ref{ghost}), we can rewrite the  identity (\ref{WTvertex}) as follows
\begin{equation}\label{WTvm}
q_\mu\Gamma^\mu(k,p,q)[1+b(q^2)] = ie [S^{-1}_{F}(k)H_1(k,p,q) - H_2(k,p,q)S^{-1}_{F}(p)].
\end{equation}
By considering that  energy-momentum conservation holds at the vertices $\Gamma^\mu(k,p,q)$ and $H(k,p,q)$, we can write
\begin{equation}
\Gamma^\mu(k,p,q) = ie(2\pi)^4 \delta^4(k-p-q)\tilde\Gamma^\mu(p,p+q)
\end{equation}
and
\begin{equation}
H(k,p,q) = (2\pi)^4 \delta^4(k-p-q)\tilde H(p,p+q). 
\end{equation}
With this representation we may obtain from Eq. (\ref{WTvm}) that
\begin{equation}\label{2}
q_\mu\tilde\Gamma^\mu(p,p+q)[1+b(q^2)] = S_{F}^{-1}(p+q)\tilde H_1(p,p+q) - \tilde H_2(p,p+q)S_{F}^{-1}(p).
\end{equation}

\subsection{The triple photon vertex}

To obtain ST identity for the triple photon vertex, the proper part of $\langle 0|T(A_\mu A_\nu A_\lambda)|0\rangle$, we differentiate the  functional equation (\ref{gfc}) with respect to $\zeta(x)$, $J^\nu(y)$ and $J^\lambda(z)$ and turn off all the sources. The result is
\begin{equation}
\frac1{\alpha e} \partial_x^\mu \frac{\delta^3 W}{\delta J^\mu(x)\delta J^\nu(y)\delta J^\lambda(z)}\left | = \frac{\delta^3 W}{\delta\zeta(x)\delta K^\nu(y)\delta J^\lambda(z)}\right | +\left. \frac{\delta^3 W}{\delta\zeta(x)\delta J^\nu(y)\delta K^\lambda(z)} \right | ,
\end{equation}
or in terms of the Green functions,
\begin{eqnarray}
-\frac1{\alpha} \partial_x^\mu \langle 0|T(A_\mu(x)A_\nu(y)A_\lambda(z))|0\rangle &=& \langle 0|T(\bar C(x)D^{AD}_\nu(y) C(y)A_\lambda(z))|0\rangle \nonumber\\
&& + \langle 0|T(\bar C(x)A_\nu(y)D^{AD}_\lambda(z)C(z))|0\rangle,
\end{eqnarray}
where $D^{AD}_\nu(y)$ denotes the covariant derivative in the adjoint representation, $D^{AD}_\nu(y) C(y)=\partial_{y\nu}C(y)-ie[A_\nu(y),C(y)]_\star$. Thus, we can rewrite the above expression  as
\begin{eqnarray}\label{triplevertex}
&&-\frac1{\alpha} \partial_{x}^{\mu} \langle 0|T(A_\mu(x)A_\nu(y)A_\lambda(z))|0\rangle = \partial_{y\nu} \langle 0|T(\bar C(x) C(y)A_\lambda(z))|0\rangle \nonumber\\
&&+ 2e\sin(\partial_y\wedge\partial_{\hat y}) \langle 0|T(\bar C(x)A_\nu(y)C(\hat y)A_\lambda(z))|0\rangle +\partial_{z\lambda} \langle 0|T(\bar C(x)A_\nu(y)C(z))|0\rangle \nonumber\\
&&+ 2e\sin(\partial_z\wedge\partial_{\hat z}) \langle 0|T(\bar C(x)A_\nu(y)A_\lambda(z)C(\hat z))|0\rangle,
\end{eqnarray}
where, after the application of the differential operators, we must identify $\hat y$ and $\hat z$ respectively with $y$ and $z$.

These Green functions have the following one-particle irreducible decomposition:
\begin{eqnarray}
&&\langle 0|T(A_\mu(x)A_\nu(y)A_\lambda(z))|0\rangle \nonumber\\
&&\quad\quad\quad\quad =  \int d^4x'd^4y'd^4z' D_{\mu\mu'}(x-x')D_{\nu\nu'}(y-y')D_{\lambda\lambda'}(z-z')\Gamma^{\mu'\nu'\lambda'}(x',y',z'), \\
&&\partial_{y\nu}\langle 0|T(\bar C(x) C(y)A_\lambda(z))|0\rangle \nonumber\\
&&\quad\quad\quad\quad = i\int d^4x'd^4y'd^4z' \Delta(x-x')\partial_{y\nu}\Delta(y-y')D_{\lambda\rho}(z-z')G^\rho(x',y',z'), \\
&& 2e\sin(\partial_y\wedge\partial_{\hat{y}}) \langle 0|T(\bar C(x)A_\nu(y)C(\hat{y})A_\lambda(z))|0\rangle \nonumber\\
&&\quad\quad\quad\quad = -i\int d^4x'd^4z' \Delta(x'-x) G_\nu^{\;\;\lambda'}(x',y,z')D_{\lambda'\lambda}(z'-z),\\
&&\partial_{z\lambda} \langle 0|T(\bar C(x)A_\nu(y)C(z))|0\rangle \nonumber\\
&&\quad\quad\quad\quad = i\int d^4x'd^4y'd^4z' \Delta(x'-x)D_{\nu\rho}(y-y')\partial_{z\lambda}\Delta(z-z')G^\rho(x',z',y'),\\
&&2e\sin(\partial_z\wedge\partial_{\hat{z}}) \langle 0|T(\bar C(x)A_\nu(y)A_\lambda(z)C(\hat{z}))|0\rangle \nonumber\\
&&\quad\quad\quad\quad = -i\int d^4x'd^4y' \Delta(x'-x) {G_\lambda}^{\nu'}(x',z,y')D_{\nu'\nu}(y'-y), 
\end{eqnarray}
where  $\Gamma^{\mu'\nu'\lambda'}(x',y',z')$ and $G^\rho(x',z',y')$ are the triple gauge and the ghost-gauge vertices respectively,
\begin{equation}\label{AAA}
\Gamma^{\mu'\nu'\lambda'}(x',y',z') = \frac{\delta^3 \Gamma}{\delta A_{cl\,\mu'}(x')\delta A_{cl\,\nu'}(y')\delta A_{cl\,\lambda'}(z')}
\end{equation}
and
\begin{equation}
G^\rho(x',z',y') = \frac{\delta^3 \Gamma}{\delta C_{cl}(x')\delta\bar C_{cl}(y')\delta A_{cl\,\rho}(z')}.
\end{equation}
Also
\begin{equation}\label{G}
{G_\nu}^{\lambda'}(x',y,z') = -2e\int d^4u\,d^4v\, \sin(\partial_y\wedge\partial_{\hat{y}}) \Delta(y-u)D_\nu^{\;\;\nu'}(\hat{y}-v)\Gamma_{\nu'}^{\;\;\;\lambda'}(x',u,v,z'), 
\end{equation}
in which 
\begin{equation}
\Gamma_{\nu'}^{\;\;\;\lambda'}(x',u,v,z') = \frac{\delta^4 \Gamma}{\delta C_{cl}(x')\delta\bar C_{cl}(u)\delta A^{\nu'}_{cl}(v)\delta A_{cl\,\lambda'}(z')}.
\end{equation}

In momentum space the Eq.~(\ref{triplevertex}) reads
\begin{eqnarray}\label{3}
p^\mu \Gamma_{\mu\nu\lambda}(p,q,k)[1+b(p^2)] &=& G_{\lambda\nu'}(p,q,k)\{(q^2g^{\nu'}_{\;\;\;\nu}-q^{\nu'}q_{\nu})[1+\Pi_\T(q^2)] + \Pi_\theta(q^2)\tilde{q}^{\nu'}\tilde{q}_\nu\} \nonumber\\
&+& \!\!\! G_{\nu\lambda'}(p,k,q)\{(k^2g^{\lambda'}_{\;\;\;\lambda}-k^{\lambda'}k_{\lambda})[1+\Pi_\T(k^2)] + \Pi_\theta(k^2)\tilde{k}^{\lambda'}\tilde{k}_\lambda\}
\end{eqnarray}
with
\begin{equation}
G_{\lambda\nu'}(p,q,k)=-2e\int \frac{d^4q'}{(2\pi)^4} \sin(q'\wedge q)i\Delta(q')iD_{\nu'}^{\;\;\;\alpha}(q-q')\Gamma_{\alpha\lambda}(p,q',q-q',k),
\end{equation}
where we have used the ghost propagator (\ref{ghost}) and the inverse of the photon propagator
\begin{equation}
iD^{-1}_{\mu\nu}(q) = -i \left\{\left(q^2g_{\mu\nu}-q_\mu q_\nu\right)[1+\Pi_\T(q^2)] - \frac{q_\mu q_\nu}{\alpha} - \Pi_\theta(q^2) \tilde{q}_\mu\tilde{q}_\nu\right\},
\end{equation}
which satisfy
\begin{equation}
q^\mu D^{-1}_{\mu\nu}(q) = -\frac{1}{\alpha} q^2 q_\nu.
\end{equation} 

Using the energy-momentum conservation at the vertices $\Gamma_{\mu\nu\lambda}(p,q,k)$ and $G_{\nu'\lambda}(p,q,k)$, so that
\begin{equation}
\Gamma_{\mu\nu\lambda}(p,q,k) = (2\pi)^4 \delta^4(p+q+k)\tilde\Gamma_{\mu\nu\lambda}(p,q,-p-q)
\end{equation}
and
\begin{equation}
G_{\lambda\nu'}(p,q,k) = (2\pi)^4 \delta^4(p+q+k) \tilde G_{\lambda\nu'}(p,q,-p-q),
\end{equation}
we get
\begin{eqnarray}\label{3}
p^\mu \tilde\Gamma_{\mu\nu\lambda}(p,q,k)[1+b(p^2)] &=& 
\tilde G_{\lambda\nu'}(p,q,k)\{(q^2{g^{\nu'}}_{\nu}-q^{\nu'}q_{\nu})[1+\Pi_\T(q^2)] + \Pi_\theta(q^2)\tilde{q}^{\nu'}\tilde{q}_\nu\} \\
&&+\tilde G_{\nu\lambda'}(p,k,q)\{[k^2{g^{\lambda'}}_{\lambda}-k^{\lambda'}k_{\lambda}][1+\Pi_\T(k^2)] + \Pi_\theta(k^2)\tilde{k}^{\lambda'}\tilde{k}_\lambda\}. \nonumber
\end{eqnarray}

\section{Slavnov-Taylor identities at one-loop: explicit calculations}

\subsection{The vectorial vertex function}

The ST identities derived previously are valid only in a formal way  since  the radiative corrections contain ultraviolet divergences.
To eliminate these  divergences counterterms must be introduced so that
the action for noncommutative QED$_4$ becomes
\begin{eqnarray}
S&=& \int d^4 x \left [-\frac{Z_3}{4} G_{\mu\nu}G^{\mu\nu}- \frac{ieZ_1}{2} [A_\mu,A_\nu]_\star G^{\mu\nu}
+ \frac{e^2 Z_4}{4} [A_\mu,A_\nu]_\star [A^\mu,A^\nu]_\star+ Z_{2}\bar\psi  i \slash \!\!\!{\partial}\psi\right. \nonumber\\
&&\!\!- (m+\delta m) \bar \psi \psi+ e\left. Z_{1F} \bar \psi\star \slash\!\!\!\! A \psi  - \frac{1}{2\alpha} (\partial_\mu A^\mu)^2 + \tilde Z_3\partial_\mu \bar C  \partial^\mu C - ie\tilde Z_1 \partial_\mu \bar C \star [A^\mu,C]_\star \right ],
\end{eqnarray}
where $G_{\mu\nu}=\partial_\mu A_\nu - \partial_\nu A_\mu$.

We begin  by considering  the one-loop contributions to the vectorial vertex function. Writing the $\tilde\Gamma^\mu(p,p+q)$, $S^{-1}(p+q)$ and $\tilde H(p,p+q)$ expansions as
\begin{equation}
\tilde\Gamma^\mu(p,p+q) = Z_{1F}\gamma^\mu e^{ip\wedge q} + \Lambda^\mu(p,p+q)e^{ip\wedge q},
\end{equation}
\begin{equation}
\tilde H_i(p,p+q) =  \left[Z_{i+4} e^{ip\wedge q} + B_i(p,p+q)e^{ip\wedge q}\right] \qquad \mbox{for $i=1,2$}
\end{equation}
and
\begin{equation}
S_{F}^{-1}(p) = \left[Z_2\slashed{p}-m-\delta m - \Sigma(p)\right],
\end{equation}
where $\Lambda^\mu(p,p+q)$, $\Sigma(p)$ and $B(p,p+q)$ are the one-loop contributions. Thus, the ST identity for the vectorial vertex (\ref{2}), in the tree approximation, becomes
\begin{eqnarray}\label{tree}
Z_{1F}\tilde Z_3\slashed{q}e^{ip\wedge q} = [Z_2 Z_5 (\slashed{p} + \slashed{q}) - (m+\delta m) Z_5]  e^{ip\wedge q} - [Z_2 Z_6 \slashed{p} - (m+\delta m) Z_6]  e^{ip\wedge q},
\end{eqnarray}
so that the validity of the ST identity requires that
\begin{equation}
 Z_5=Z_6\qquad \mbox{and}\qquad \tilde Z_3/Z_5= Z_2/Z_{1F} .
\end{equation}
For the one-loop approximation, we have
\begin{eqnarray}\label{loop}
q_\mu \Lambda^\mu_a(p,p+q) + q_\mu \Lambda^\mu_b(p,p+q) + \slashed q b(q^2) &=& \Sigma(p) - \Sigma(p+q) + (\slashed{p}+\slashed{q}-m)B_1(p,p+q) \nonumber\\
&& - B_2(p,p+q)(\slashed{p}-m).
\end{eqnarray}
 
The diagrams  representing these contributions are given by: (from now on, we restrict ourselves to the Feynman gauge, $\alpha=1$)
\begin{eqnarray}
1.\raisebox{-0.4cm}{\incps{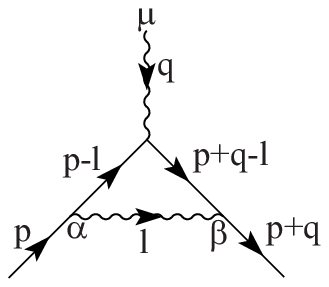}{-1.5cm}{-0.5cm}{1.5cm}{1.5cm}}
 &=& -ie \Lambda^\mu_a(p,p+q) e^{i p\wedge q} \\ \nonumber\\ \nonumber\\
 &=&  \int \frac{d^4l}{(2\pi)^4} \frac{-ig_{\alpha\beta}}{l^2} (-ie\gamma^\alpha) iS_0(p+q-l) (-ie\gamma^\mu) iS_0(p-l) (-ie\gamma^\beta) e^{-2i l\wedge q}e^{i p\wedge q}, \nonumber
\end{eqnarray}
with
\begin{equation}
iS_0(p)=\frac{i}{\slashed{p}-m}.
\end{equation}
This contribution is entirely nonplanar and using $ \slashed{q}= (\slashed{q}+\slashed p-\slashed l-m)-(\slashed p-\slashed l-m)$ can be shown to satisfy
\begin{equation}
q_\mu \Lambda^\mu_a(p,p+q) = \Sigma_\np(p) - \Sigma_\np(p+q),
\end{equation}
where the nonplanar fermion self-energy is
\begin{equation}\label{sigma}
-i\Sigma_\np(p) = \int \frac{d^4l}{(2\pi)^4} \frac{-ig_{\alpha\beta}}{l^2}(-ie\gamma^\alpha)iS_0(p-l)(-ie\gamma^\beta)e^{-2il\wedge q}.
\end{equation}
\begin{eqnarray}\label{L2}
2.\raisebox{-0.4cm}{\incps{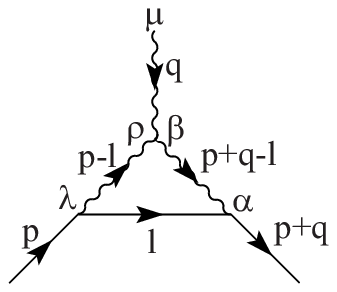}{-1.5cm}{-0.5cm}{1.5cm}{1.5cm}}
 &=& -ie \Lambda^\mu_b(p,p+q) e^{i p\wedge q} \\ \nonumber\\ \nonumber\\
 &=& \int \frac{d^4l}{(2\pi)^4} \frac{-ig_{\alpha\beta}}{(p+q-l)^2}\frac{-ig_{\lambda\rho}}{(p-l)^2} (-ie\gamma^\alpha) iS_0(l) (-ie\gamma^\lambda) \nonumber\\
 && \times (2e)\gamma^{\mu\beta\rho}(q,-p-q+l,p-l)\frac1{2i}\left(1-e^{2i l\wedge q}e^{-2i p\wedge q}\right)e^{i p\wedge q}, \nonumber
\end{eqnarray}
whose planar part logarithimically diverges. In fact its pole part (PP) is given by
\begin{equation}
 \mathrm{PP}[-ie \Lambda^\mu_b(p,p+q)] = \frac{3}{16\pi^{2}}\frac{1}{\epsilon}\gamma^\mu
\end{equation}
\noindent
so that, in the minimal dimensional regularization scheme, 
\begin{equation}
Z_{1F} = 1- \frac{3}{16\pi^{2}}\frac{1}{\epsilon}, 
\end{equation}
which agrees with the result of previous calculation \cite{Hayakawa}. Contracting $q_\mu$ in the expression~(\ref{L2}), we get
\begin{eqnarray}
q_\mu \Lambda^\mu_b(p,p+q) &=&  \int \frac{d^4l}{(2\pi)^4} \frac{-ig_{\alpha\beta}}{(p+q-l)^2}\frac{-ig_{\lambda\rho}}{(p-l)^2} (-ie\gamma^\alpha) iS_0(l) (-ie\gamma^\lambda) \left(1-e^{2i l\wedge q}e^{-2i p\wedge q}\right) \nonumber\\
&& \times \left\{(p+q-l)^\rho q^\beta + (p-l)^\beta q^\rho - [(p+q-l)\cdot q  + (p-l)\cdot q] g^{\beta\rho}\right\},
\end{eqnarray}
which may be further simplified using
\begin{eqnarray}
(p+q-l)^\rho q^\beta + (p-l)^\beta q^\rho &=& (p+q-l)^\rho (p+q-l)^\beta - (p-l)^\rho (p-l)^\beta,\;\mathrm{and} \nonumber\\
(p+q-l)\cdot q  + (p-l)\cdot q &=& (p+q-l)^2 - (p-l)^2,
\end{eqnarray}
to yield
\begin{eqnarray}
q_\mu \Lambda^\mu_b(p,p+q) &=& \Sigma(p) - \Sigma(p+q) - \Sigma_\np(p) + \Sigma_\np(p+q) \\
&& + \int \frac{d^4l}{(2\pi)^4} \frac{i}{(p+q-l)^2}\frac{i}{(p-l)^2}(ie)(\slashed{p}+\slashed{q}-\slashed{l})iS_0(l)(ie)(\slashed{p}+\slashed{q}-\slashed{l})\left(1-e^{2i l\wedge q}e^{-2i p\wedge q}\right) \nonumber\\
&& - \int \frac{d^4l}{(2\pi)^4} \frac{i}{(p+q-l)^2}\frac{i}{(p-l)^2}(ie)(\slashed{p}-\slashed{l})iS_0(l)(ie)(\slashed{p}-\slashed{l})\left(1-e^{2i l\wedge q}e^{-2i p\wedge q}\right), \nonumber
\end{eqnarray}
where $\Sigma$ is similar to $\Sigma_\np$ of Eq.~(\ref{sigma}), but without the phase factor. 

For the last term in left-hand side of Eq.~(\ref{loop}), we get
\begin{eqnarray}\label{gse}
\raisebox{-0.4cm}{\incps{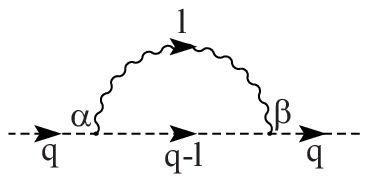}{-1.5cm}{-0.5cm}{2.0cm}{1.5cm}}
 &=& i q^2 b(q^2) \\ \nonumber\\ \nonumber\\
 &=& - \int \frac{d^4l}{(2\pi)^4} \frac{-ig_{\alpha\beta}}{l^2} \frac{i}{(q-l)^2} (2e)(q-l)^\alpha(2e)q^\beta \frac12 \left(1- e^{2il\wedge q}\right). \nonumber
\end{eqnarray}
Then,
\begin{eqnarray}
\slashed{q}b(q^2) &=& \frac{2i}{q^2} \int \frac{d^4l}{(2\pi)^4} \frac{i}{l^2}\frac{i}{(q-l)^2}(ie)\slashed{q}(ie)(q-l)\cdot q \left(1 - e^{2il\wedge q}\right) \\
&\overbrace{=}^{l \rightarrow l-p}& \frac{2i}{q^2} \int \frac{d^4l}{(2\pi)^4} \frac{i}{(p-l)^2}\frac{i}{(p+q-l)^2}(ie)\slashed{q}(ie)(p+q-l)\cdot q \left(1 - e^{2il\wedge q}e^{-2ip\wedge q}\right). \nonumber
\end{eqnarray}
Since
\begin{equation}
2(p+q-l)\cdot q = (p+q-l)^2 - (p-l)^2 + q^2,
\end{equation}
we find
\begin{equation}
\slashed{q}b(q^2) = i \int \frac{d^4l}{(2\pi)^4} \frac{i}{(p-l)^2}\frac{i}{(p+q-l)^2}(ie)\slashed{q}(ie)\left(1 - e^{2il\wedge q}e^{-2ip\wedge q}\right).
\end{equation}

Hence, for the left-hand side of Eq. (\ref{loop}), we find
\begin{eqnarray}\label{rhs}
&& q_\mu \Lambda^\mu_a(p,p+q) + q_\mu \Lambda^\mu_b(p,p+q)\; + \slashed{q}b(q^2) = \Sigma(p) - \Sigma(p+q) \nonumber\\
&& + \int \frac{d^4l}{(2\pi)^4} \frac{i}{(p+q-l)^2}\frac{i}{(p-l)^2}(ie)(\slashed{p}+\slashed{q}-\slashed{l})iS_0(l)(ie)(\slashed{p}+\slashed{q}-\slashed{l})\left(1-e^{2i l\wedge q}e^{-2i p\wedge q}\right) \nonumber\\
&& - \int \frac{d^4l}{(2\pi)^4} \frac{i}{(p+q-l)^2}\frac{i}{(p-l)^2}(ie)(\slashed{p}-\slashed{l})iS_0(l)(ie)(\slashed{p}-\slashed{l})\left(1-e^{2i l\wedge q}e^{-2i p\wedge q}\right) \nonumber\\
&& + i \int \frac{d^4l}{(2\pi)^4} \frac{i}{(p-l)^2}\frac{i}{(p+q-l)^2}(ie)\slashed{q}(ie)\left(1 - e^{2il\wedge q}e^{-2ip\wedge q}\right).
\end{eqnarray}

Let us now consider the one-loop contributions to the right-hand side of Eq. (\ref{loop}). We have
\begin{eqnarray}
\raisebox{-0.4cm}{\incps{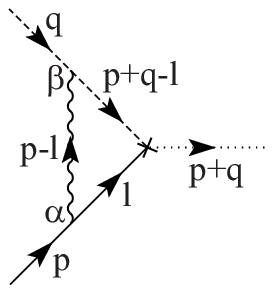}{-1.5cm}{-0.5cm}{1.2cm}{1.5cm}}
 &=& B_1(p,p+q)e^{ip\wedge q} \\ \nonumber\\ \nonumber\\
 &=&  -\int \frac{d^4l}{(2\pi)^4} iS_0(l)\frac{i}{(p+q-l)^2}(-ie\gamma^\alpha)\frac{-ig_{\alpha\beta}}{(p-l)^2}(2e)(p+q-l)^\beta \nonumber\\ 
&& \times \frac{1}{2i}\left(1 - e^{2il\times q}e^{-2ip\times q}\right)e^{ip\times q}, \nonumber
\end{eqnarray}
yielding 	
\begin{equation}
 Z_5= 1-\frac{e^2}{16\pi^2} \frac{1}{\epsilon}.
\end{equation}
Notice that  in the Landau gauge $Z_5=1$ as it happens in ordinary commutative QCD \cite{marciano}.  Now,
\begin{eqnarray}
(\slashed{p}\;+\slashed{q}-m)B_1(p,p+q) &=& \int \frac{d^4l}{(2\pi)^4} \frac{i}{(p+q-l)^2}\frac{i}{(p-l)^2}(ie)(\slashed{p}+\slashed{q}-m)iS_0(l)(ie)(\slashed{p}+\slashed{q}-\slashed{l})\nonumber\\
&&\times\left(1-e^{2i l\times q}e^{-2i p\times q}\right)
\end{eqnarray}
and making the substitution $(\slashed{p}+\slashed{q}-m) \rightarrow (\slashed{p}+\slashed{q}-\slashed{l}) + (\slashed{l} - m)$, we obtain
\begin{eqnarray}
(\slashed{p}\;+\slashed{q}-m)B_1(p,p+q) &=& \int \frac{d^4l}{(2\pi)^4} \frac{i}{(p+q-l)^2}\frac{i}{(p-l)^2}(ie)(\slashed{p}+\slashed{q}-\slashed{l})iS_0(l)(ie)(\slashed{p}+\slashed{q}-\slashed{l})\nonumber\\
&&\times\left(1-e^{2i l\wedge q}e^{-2i p\wedge q}\right) \nonumber\\
&& + i \int \frac{d^4l}{(2\pi)^4} \frac{i}{(p+q-l)^2}\frac{i}{(p-l)^2}(ie)(\slashed{p}+\slashed{q}-\slashed{l})(ie)\nonumber\\
&&\times\left(1-e^{2il\wedge q}e^{-2ip\wedge q}\right).
\end{eqnarray}
Similarly,
\begin{eqnarray}
\raisebox{-0.4cm}{\incps{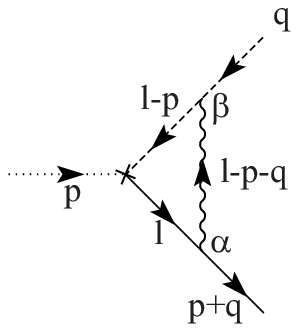}{-1.5cm}{-0.5cm}{1.7cm}{1.5cm}}
 &=& B_2(p,p+q)e^{ip\wedge q} \\ \nonumber\\ \nonumber\\
 &=& \int \frac{d^4l}{(2\pi)^4} (-ie\gamma^\alpha) \frac{-ig_{\alpha\beta}}{(p+q-l)^2}(2e)(l-p)^\beta  \frac{i}{(p-l)^2} iS_0(l) \nonumber\\ 
 &&\times\frac{1}{2i}\left(1 - e^{2il\times q}e^{-2ip\times q}\right)e^{ip\times q}. \nonumber
\end{eqnarray}
Therefore,
\begin{eqnarray}
B_2(p,p+q)(\slashed{p}-m) &=& \int \frac{d^4l}{(2\pi)^4} \frac{i}{(p+q-l)^2}\frac{i}{(p-l)^2}(ie)(\slashed{p}-\slashed{l})iS_0(l)(ie)(\slashed{p}-\slashed{l}) \nonumber\\
&&\times\left(1-e^{2il\wedge q}e^{-2ip\wedge q}\right) \nonumber\\
&& + i \int \frac{d^4l}{(2\pi)^4} \frac{i}{(p+q-l)^2}\frac{i}{(p-l)^2} (ie)(\slashed{p}-\slashed{l})(ie) \nonumber\\
&&\times\left(1-e^{2il\times q}e^{-2ip\times q}\right),
\end{eqnarray}
where we also have done the replacement $(\slashed{p}-m) \rightarrow (\slashed{p}-\slashed{l}) + (\slashed{l} - m)$. Finally, summing the above results, 
\begin{eqnarray}
&&(\slashed{p}+\slashed{q}-m)B_1(p,p+q) - B_2(p,p+q)(\slashed{p}-m) = \nonumber\\
&& + \int \frac{d^4l}{(2\pi)^4} \frac{i}{(p+q-l)^2}\frac{i}{(p-l)^2}(ie)(\slashed{p}+\slashed{q}-\slashed{l})iS_0(l)(ie)(\slashed{p}+\slashed{q}-\slashed{l})\left(1-e^{2i l\wedge q}e^{-2i p\wedge q}\right) \nonumber\\
&& - \int \frac{d^4l}{(2\pi)^4} \frac{i}{(p+q-l)^2}\frac{i}{(p-l)^2}(ie)(\slashed{p}-\slashed{l})iS_0(l)(ie)(\slashed{p}-\slashed{l})\left(1-e^{2i l\wedge q}e^{-2i p\wedge q}\right) \nonumber\\
&& + i \int \frac{d^4l}{(2\pi)^4} \frac{i}{(p-l)^2}\frac{i}{(p+q-l)^2}(ie)\slashed{q}(ie)\left(1 - e^{2il\wedge q}e^{-2ip\wedge q}\right).
\end{eqnarray}
Thus, the left-hand side of Eq.~(\ref{loop}) is identical to the right-hand side as we can see from Eq.~(\ref{rhs}). Therefore, the ST identity for the vectorial vertex is satisfied at one-loop.

\subsection{The triple photon vertex}

Writing the expansions for $\tilde\Gamma^{\mu\nu\lambda}(p,q,k)$ and $\tilde G^\lambda_{\;\;\nu'}(p,q,k)$, defined in~(\ref{AAA}) and~(\ref{G}), as
\begin{equation}
\tilde\Gamma^{\mu\nu\lambda}(p,q,k) = 2eZ_1\sin(p\wedge q)\gamma^{\mu\nu\lambda}(p,q,k) + 2e\sin(p\wedge q)\Lambda^{\mu\nu\lambda}(p,q,k),
\end{equation}
\begin{equation}
\tilde G^\lambda_{\;\;\nu'}(p,q,k) = -2e\tilde Z_1\sin(p\wedge q){g^\lambda}_{\nu'} + 2e\sin(p\wedge q){B^\lambda}_{\nu'}(p,q,k),
\end{equation}
we obtain the ST identity (\ref{3}) for the triple photon vertex, in the tree approximation, 
\begin{equation}
Z_1\tilde Z_3\left[(k^2 g^{\lambda\nu} - k^\lambda k^\nu)-(q^2 g^{\nu\lambda} - q^\nu q^\lambda)\right]=\tilde Z_1Z_3\left[(k^2 g^{\lambda\nu} - k^\lambda k^\nu)-(q^2 g^{\nu\lambda} - q^\nu q^\lambda)\right],
\end{equation}
which requires that
\begin{equation}
\tilde Z_3/\tilde Z_1=Z_3/Z_1.
\end{equation}
On the other hand, the one-loop approximation is given by
\begin{eqnarray}\label{ITF}
&& p_\mu \gamma^{\mu\nu\lambda}(p,q,k)b(p^2) + p_\mu \Lambda^{\mu\nu\lambda}(p,q,k) = \Pi^{\nu\lambda}(q) - \Pi^{\lambda\nu}(k) \nonumber\\
&&+ B^\lambda_{\;\;\nu'}(p,q,k)(q^2 g^{\nu\nu'} - q^\nu q^{\nu'}) + B^\nu_{\;\;\lambda'}(p,k,q)(k^2 g^{\lambda\lambda'} - k^\lambda k^{\lambda'}),
\end{eqnarray}
where we have introduced the photon self-energy $\Pi^{\nu\nu'}(q)\equiv\Pi_\T(q^2)(q^2g^{\nu\nu'}-q^{\nu}q^{\nu'}) + \Pi_\theta(q^2)\tilde{q}^{\nu}\tilde{q}^{\nu'}$.

The contributions with a fermion loop in the left and right-hand sides of Eq.~(\ref{ITF}) are directly identified when we consider the diagrams:
\begin{eqnarray}
\raisebox{-0.4cm}{\incps{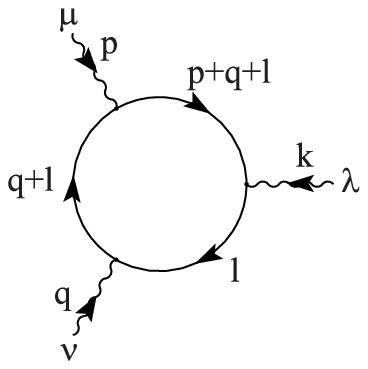}{-1.8cm}{-0.5cm}{2cm}{1.8cm}} &=& 2e\sin(p\wedge q)\Lambda_{a1}^{\mu\nu\lambda}(p,q,k) \\ \nonumber\\ \nonumber\\
 &=& -\int \frac{d^4l}{(2\pi)^4} \tr(-ie\gamma^\nu)iS_0(q+l)(-ie\gamma^\mu)iS_0(p+q+l)\nonumber\\
 &&\times(-ie\gamma^\lambda)iS_0(l)e^{-i p\wedge q},  \nonumber \\
\raisebox{-0.4cm}{\incps{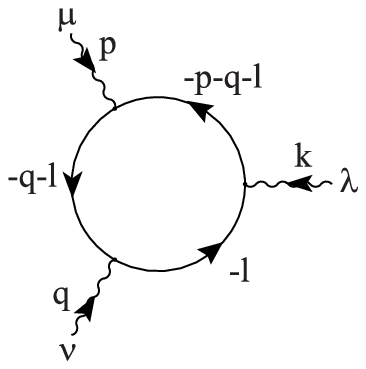}{-1.8cm}{-0.5cm}{2cm}{1.8cm}} &=&  2e\sin(p\wedge q)\Lambda_{a2}^{\mu\nu\lambda}(p,q,k) \\ \nonumber\\ \nonumber\\
 &=& -\int \frac{d^4l}{(2\pi)^4} \tr\,iS_0(-l)(-ie\gamma^\lambda)iS_0(-p-q-l)\nonumber\\
 &&\times(-ie\gamma^\mu)iS_0(-q-l)(-ie\gamma^\nu)e^{i p\wedge q}.  \nonumber
\end{eqnarray}
These diagrams are different only in the circulation of the momentum integration. Since $C\gamma^\mu C^{-1}=-\gamma^{\T\mu}$ and $CS_0(l)C^{-1}=S_0^\T(-l)$, we can rewrite the above expression as
\begin{eqnarray}
2e\sin(p\wedge q)\Lambda_{a2}^{\mu\nu\lambda}(p,q,k) 
 &=& e^3 \int \frac{d^4l}{(2\pi)^4} \tr\,S_0^\T(l)\gamma^{\T\lambda}S_0^\T(p+q+l)\gamma^{\T\mu}S_0(q+l)\gamma^{\T\nu} e^{i p\wedge q}  \nonumber\\
 &=& e^3 \int \frac{d^4l}{(2\pi)^4} \tr[\gamma^\nu S_0(q+l)\gamma^\mu S_0(p+q+l)\gamma^\lambda S_0(l)]^\T e^{i p\wedge q} \nonumber\\
 &=& e^3 \int \frac{d^4l}{(2\pi)^4} \tr\,\gamma^\nu S_0(q+l)\gamma^\mu S_0(p+q+l)\gamma^\lambda S_0(l) e^{i p\wedge q}.
\end{eqnarray}
Thus, by summing the two diagrams, we obtain
\begin{eqnarray}
2e\sin(p\wedge q)\Lambda_{a}^{\mu\nu\lambda}(p,q,k) &=& 2e \sin(p\wedge q)\left[\Lambda_{a1}^{\mu\nu\lambda}(p,q,k)+\Lambda_{a2}^{\mu\nu\lambda}(p,q,k)\right] \\ 
 &=& 2ie^3 \int \frac{d^4l}{(2\pi)^4} \tr\,\gamma^\nu S_0(q+l)\gamma^\mu S_0(p+q+l)\gamma^\lambda S_0(l) \sin(p\wedge q), \nonumber
\end{eqnarray}
i.e.,
\begin{equation}
\Lambda_{a}^{\mu\nu\lambda}(p,q,k) = ie^2 \int \frac{d^4l}{(2\pi)^4} \tr\,\gamma^\nu S_0(q+l)\gamma^\mu S_0(p+q+l)\gamma^\lambda S_0(l).
\end{equation}
Therefore, using also $(\slashed{p}+\slashed{q}+\slashed{l}-m)-(\slashed{q}+\slashed{l}-m)$, we get
\begin{equation}
p_\mu\Lambda_{a}^{\mu\nu\lambda}(p,q,k) = \Pi^{\nu\lambda}_a(q) - \Pi^{\lambda\nu}_a(k),
\end{equation}
where 
\begin{equation}
i\Pi^{\nu\lambda}_a(q) =  -\int \frac{d^4l}{(2\pi)^4} \tr(-ie\gamma^\nu)iS_0(q+l)(-ie\gamma^\lambda)iS_0(l)
\end{equation}
is the photon self-energy, with a fermion loop.

From now on, differently for the previous calculations, the contributions to the ST identity turns out to be very involved and a complete verification is unfeasible. In this situation we restrict ourselves in to verify the matching of the divergent parts of the two sides of Eq.~(\ref{ITF}).

The diagram for the ghost self-energy has already been considered in (\ref{gse}) and its PP is given by
\begin{equation}
\mathrm{PP}[p_\mu \gamma^{\mu\nu\lambda}(p,q,k)b(p^2)] = \frac{e^2}{16\pi^2}\frac{1}{\epsilon}[(k^2 g^{\lambda\nu} - k^\lambda k^\nu) - (q^2 g^{\nu\lambda} - q^\nu q^\lambda)].
\end{equation}

In the sequel we consider the diagrams $\Lambda^{\mu\nu\lambda}$ of the left-hand side of~(\ref{ITF}), with ghost loop,
\begin{eqnarray}
\raisebox{-0.4cm}{\incps{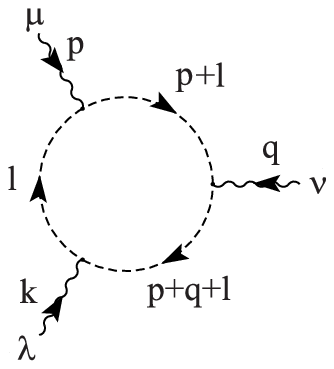}{-1.8cm}{-0.5cm}{2cm}{1.8cm}} &=& 2e\sin(p\wedge q)\Lambda_{b1}^{\mu\nu\lambda}(p,q,k) \\ \nonumber\\ \nonumber\\
 &=& -i^3(2e)^3 \int \frac{d^4l}{(2\pi)^4} \frac{l^\lambda(p+l)^\mu(p+q+l)^\nu}{l^2(p+l)^2(p+q+l)^2} \nonumber\\\
&&\times \sin(l\wedge p)\sin(l\wedge p + l\wedge q)\sin(l\wedge q + p\wedge q), \nonumber\\
\raisebox{-0.4cm}{\incps{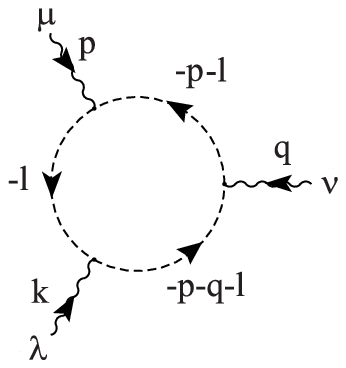}{-1.8cm}{-0.5cm}{2cm}{1.8cm}}  &=& 2e\sin(p\wedge q)\Lambda_{b2}^{\mu\nu\lambda}(p,q,k) \\ \nonumber\\ \nonumber\\
 &=& -i^3(2e)^3 \int \frac{d^4l}{(2\pi)^4} \frac{(-p-q-l)^\lambda(-p-l)^\nu(-l)^\mu}{l^2(p+l)^2(p+q+l)^2} \nonumber\\
 &&\times \sin(l\wedge p)\sin(-l\wedge p - l\wedge q)\sin(l\wedge q + p\wedge q), \nonumber
\end{eqnarray}
and photon loop,
\begin{eqnarray}
\raisebox{-0.4cm}{\incps{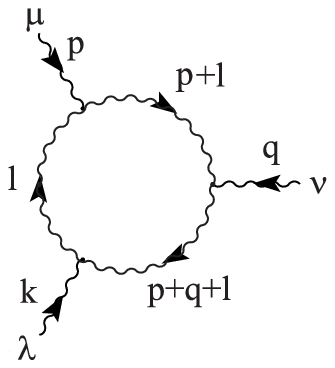}{-1.8cm}{-0.5cm}{2cm}{1.8cm}} &=& 2e\sin(p\wedge q)\Lambda_{c}^{\mu\nu\lambda}(p,q,k) \\ \nonumber\\ \nonumber\\
 &=& i^3(2e)^3 \int \frac{d^4l}{(2\pi)^4} \frac{\gamma^{\alpha\lambda\beta}(p+q+l,-p-q,-l)\gamma^{\beta\mu\rho}(l,p,-p-l)}{l^2(p+l)^2(p+q+l)^2} \nonumber\\
 && \times \gamma^{\rho\nu\alpha}(l+p,q,-p-q-l)\sin(l\wedge p)\sin(-l\wedge p - l\wedge q)\sin(l\wedge q + p\wedge q), \nonumber
\end{eqnarray}
that have the same phase factors and, therefore, can be calculated analogously. Their PP contributions are 
\begin{eqnarray}
{\rm PP}[p_\mu\Lambda_{b1,b2,c}^{\mu\nu\lambda}(p,q,k)]&=& -\frac{19e^2}{96\pi^2}\frac{1}{\epsilon}[(k^2 g^{\lambda\nu} - k^\lambda k^\nu) - (q^2 g^{\nu\lambda} - q^\nu q^\lambda)]. 
\end{eqnarray}

The remain diagrams for $\Lambda^{\mu\nu\lambda}$, are given by
\begin{eqnarray}
\raisebox{-0.4cm}{\incps{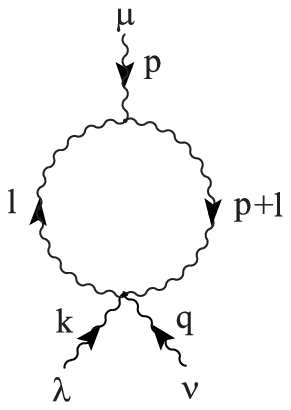}{-1.8cm}{-0.5cm}{1.6cm}{2cm}}  &=& 2e\sin(p\wedge q)\Lambda_{d1}^{\mu\nu\lambda}(p,q,k) \\ \nonumber\\ \nonumber\\
 &=& \frac12 (-i)^2(2e)(-4ie^2) \int \frac{d^4l}{(2\pi)^4} \frac{(l-p)^\lambda g^{\mu\nu}+({g^\alpha}_\alpha-2)(2l+p)^\mu g^{\nu\lambda}+(l+2p)^\nu g^{\mu\lambda}}{l^2(p+l)^2} \nonumber\\
 && \times \sin(l\wedge p)\sin(-l\wedge p - l\wedge q)\sin(l\wedge q + p\wedge q) \nonumber\\
 && + \frac12 (-i)^2(2e)(-4ie^2) \int \frac{d^4l}{(2\pi)^4} \frac{(l+2p)^\lambda g^{\mu\nu}+({g^\alpha}_\alpha-2)(2l+p)^\mu g^{\nu\lambda}+(l-p)^\nu g^{\mu\lambda}}{l^2(p+l)^2} \nonumber\\
 && \times \sin(l\wedge p)\sin(l\wedge q)\sin(-l\wedge p -l\wedge q - p\wedge q) \nonumber\\
 && + \frac12 (-i)^2(2e)(-4ie^2) \int \frac{d^4l}{(2\pi)^4} \frac{3p^\lambda g^{\mu\nu}-3p^\nu g^{\mu\lambda}}{l^2(p+l)^2} \sin^2(l\wedge p) \sin(p\wedge q), \nonumber\\
\raisebox{-0.4cm}{\incps{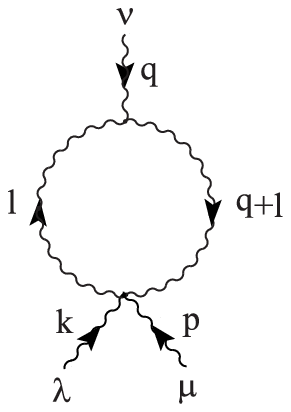}{-1.8cm}{-0.5cm}{1.6cm}{2cm}} &=& 2e\sin(p\wedge q)\Lambda_{d2}^{\mu\nu\lambda}(p,q,k)  \\ \nonumber\\ \nonumber\\
 &=& \frac12 (-i)^2(2e)(-4ie^2) \int \frac{d^4l}{(2\pi)^4} \frac{(l-q)^\lambda g^{\mu\nu}+(l+2q)^\mu g^{\nu\lambda}+({g^\alpha}_\alpha-2)(2l+q)^\nu g^{\mu\lambda}}{l^2(q+l)^2} \nonumber\\
 && \times \sin(l\wedge q)\sin(-l\wedge p - l\wedge q)\sin(l\wedge p - p\wedge q) \nonumber\\
 && + \frac12 (-i)^2(2e)(-4ie^2) \int \frac{d^4l}{(2\pi)^4} \frac{(l+2q)^\lambda g^{\mu\nu}+(l-q)^\mu g^{\nu\lambda}+({g^\alpha}_\alpha-2)(2l+q)^\nu g^{\mu\lambda}}{l^2(q+l)^2} \nonumber\\
 && \times \sin(l\wedge p)\sin(l\wedge q)\sin(-l\wedge p -l\wedge q + p\wedge q) \nonumber\\
 && + \frac12 (-i)^2(2e)(-4ie^2) \int \frac{d^4l}{(2\pi)^4} \frac{3q^\mu g^{\nu\lambda}-3q^\lambda g^{\mu\nu}}{l^2(q+l)^2} \sin^2(l\wedge q) \sin(p\wedge q),\quad\mathrm{and} \nonumber\\
\raisebox{-0.4cm}{\incps{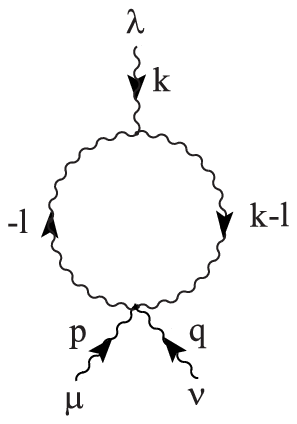}{-1.8cm}{-0.5cm}{1.6cm}{2cm}} &=& 2e\sin(p\wedge q)\Lambda_{d3}^{\mu\nu\lambda}(p,q,k)  \\ \nonumber\\ \nonumber\\
 &=& \frac12 (-i)^2(2e)(-4ie^2) \int \frac{d^4l}{(2\pi)^4} \frac{-({g^\alpha}_\alpha-2)(2l-k)^\lambda g^{\mu\nu}-(l+k)^\mu g^{\nu\lambda}-(l-2k)^\nu g^{\mu\lambda}}{l^2(p+q+l)^2} \nonumber\\
 && \times \sin(l\wedge p)\sin(-l\wedge p - l\wedge q)\sin(-l\wedge q - p\wedge q) \nonumber\\
 && + \frac12 (-i)^2(2e)(-4ie^2) \int \frac{d^4l}{(2\pi)^4} \frac{-({g^\alpha}_\alpha-2)(2l-k)^\lambda g^{\mu\nu}-(l-2k)^\mu g^{\nu\lambda}-(l+k)^\nu g^{\mu\lambda}}{l^2(p+q+l)^2} \nonumber\\
 && \times \sin(l\wedge q)\sin(-l\wedge p-l\wedge q)\sin(-l\wedge p + p\wedge q) \nonumber\\
 && + \frac12 (-i)^2(2e)(-4ie^2) \int \frac{d^4l}{(2\pi)^4} \frac{3k^\nu g^{\mu\lambda}-3k^\mu g^{\lambda\nu}}{l^2(p+q+l)^2} \sin^2(l\wedge p + l\wedge q) \sin(p\wedge q), \nonumber
\end{eqnarray}
where their PP contributions take the form
\begin{eqnarray}
{\rm PP}[p_\mu\Lambda_{d1,d2,d3}^{\mu\nu\lambda}(p,q,k)]&=& \frac{9e^2}{32\pi^2}\frac{1}{\epsilon}[(k^2 g^{\lambda\nu} - k^\lambda k^\nu) - (q^2 g^{\nu\lambda} - q^\nu q^\lambda)].
\end{eqnarray}
Therefore, the sum of all PP contributions of the left-hand side of Eq.~(\ref{ITF}) becomes
\begin{eqnarray}\label{sumL}
&& {\rm PP}[p_\mu(\gamma^{\mu\nu\lambda}(p,q,k)b(p^2)+\Lambda_{b1,b2,c,d1,d2,d3}^{\mu\nu\lambda}(p,q,k))] \nonumber\\ 
&&\quad\quad\quad\quad\quad\quad = \frac{7e^2}{48\pi^2}\frac{1}{\epsilon}[(k^2 g^{\lambda\nu} - k^\lambda k^\nu) - (q^2 g^{\nu\lambda} - q^\nu q^\lambda)].
\end{eqnarray}

Let us now look to the right-hand side of the identity~(\ref{ITF}). The diagrams are
\begin{eqnarray}
\raisebox{-0.4cm}{\incps{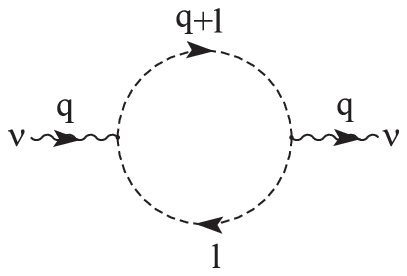}{-1.8cm}{-0.5cm}{2cm}{1.5cm}} &=& i\Pi^{\nu\lambda}_b(q) \\ \nonumber\\ \nonumber\\
 &=& -i^2(2e)^2 \int \frac{d^4l}{(2\pi)^4} \frac{l^\lambda(l+q)^\nu}{l^2(q+l)^2} \sin^2(l\wedge q), \nonumber\\
\raisebox{-0.4cm}{\incps{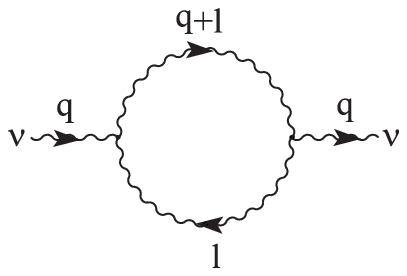}{-1.8cm}{-0.5cm}{2cm}{1.5cm}} &=& i\Pi^{\nu\lambda}_c(q) \\ \nonumber\\ \nonumber\\
 &=& \frac12(-i)^2(2e)^2 \int \frac{d^4l}{(2\pi)^4} \frac{\gamma^{\alpha\nu\beta}(l,q,-l-q)\gamma^{\beta\nu\alpha}(l+q,-q,-l)}{l^2(q+l)^2} \sin^2(l\wedge q), \nonumber\\
\raisebox{-1.2cm}{\incps{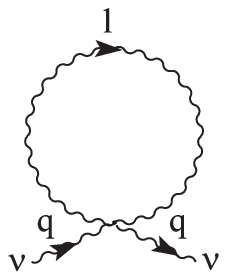}{-1.8cm}{-0.5cm}{1.5cm}{2.5cm}} &=& i\Pi^{\nu\lambda}_d(q) \\ \nonumber\\ \nonumber\\
 &=& \frac12(-i)(-4ie^2) \int \frac{d^4l}{(2\pi)^4} \frac{{2(g^\alpha}_\alpha-1)g^{\nu\lambda}}{l^2} \sin^2(l\wedge q), \nonumber
\end{eqnarray}
so that
\begin{eqnarray}
{\rm PP}[\Pi^{\nu\lambda}_{b,c,d}(q) - \Pi^{\lambda\nu}_{b,c,d}(k)] &=& \frac{5e^2}{24\pi^2}\frac{1}{\epsilon}[(k^2 g^{\lambda\nu} - k^\lambda k^\nu) - (q^2 g^{\nu\lambda} - q^\nu q^\lambda)].
\end{eqnarray}
Finally, let us see the diagrams $B^\lambda_{\;\;\nu'}(p,q,k)$ and $B^\nu_{\;\;\lambda'}(p,k,q)$, given by
\begin{eqnarray}
\raisebox{-0.4cm}{\incps{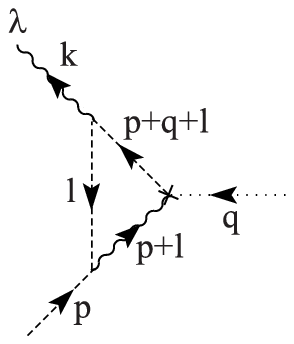}{-1.8cm}{-0.5cm}{1.2cm}{2cm}} &=&2e\sin(p\wedge q)B^\lambda_{a\,\nu'}(p,q,k) \\ \nonumber\\ \nonumber\\
 &=& i^2(-i)(2e)^3 \int \frac{d^4l}{(2\pi)^4}\frac{(-l)_{\nu'}(-p-q-l)^\lambda}{l^2(p+l)^2(p+q+l)^2}\nonumber\\
 &&\times\sin(l\wedge p)\sin(-l\wedge p - l\wedge q)\sin(-l\wedge q - p\wedge q), \nonumber\\
\raisebox{-0.4cm}{\incps{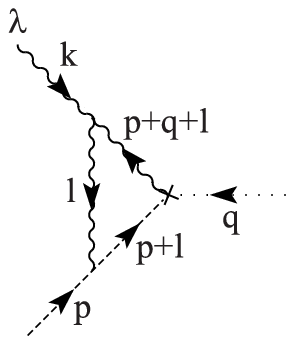}{-1.8cm}{-0.5cm}{1.2cm}{2cm}} &=&2e\sin(p\wedge q)B^\lambda_{b\,\nu'}(p,q,k) \\ \nonumber\\ \nonumber\\
 &=& i(-i)^2(2e)^3 \int \frac{d^4l}{(2\pi)^4}\frac{(p+l)_{\alpha}{\gamma^{\alpha\lambda}}_{\nu'}(-l,-p-q,p+q+l)}{l^2(p+l)^2(p+q+l)^2}\nonumber\\
 &&\times\sin(l\wedge p)\sin(-l\wedge p - l\wedge q)\sin(l\wedge q + p\wedge q), \nonumber\\
\raisebox{-0.4cm}{\incps{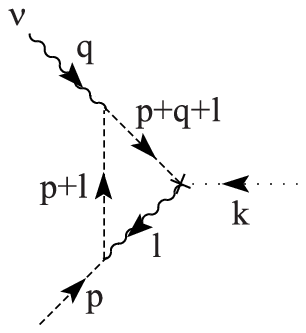}{-1.8cm}{-0.5cm}{1.2cm}{2cm}} &=&2e\sin(p\wedge q)B^\nu_{a\,\lambda'}(p,k,q) \\ \nonumber\\ \nonumber\\
 &=& i^2(-i)(2e)^3 \int \frac{d^4l}{(2\pi)^4}\frac{(p+l)_{\lambda'}(p+q+l)^\nu}{l^2(p+l)^2(p+q+l)^2}\nonumber\\
 &&\times\sin(l\wedge p)\sin(l\wedge p + l\wedge q)\sin(l\wedge q + p\wedge q),\quad\mathrm{and} \nonumber\\
\raisebox{-0.4cm}{\incps{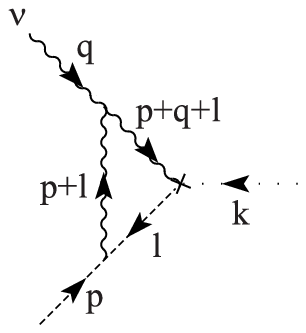}{-1.8cm}{-0.5cm}{1.2cm}{2cm}} &=&2e\sin(p\wedge q)B^\nu_{b\,\lambda'}(p,q,k) \\ \nonumber\\ \nonumber\\
 &=& i(-i)^2(2e)^3 \int \frac{d^4l}{(2\pi)^4}\frac{(-l)_{\alpha}{\gamma^{\alpha\nu}}_{\lambda'}(p+l,q,-p-q-l)}{l^2(p+l)^2(p+q+l)^2}\nonumber\\
 &&\times\sin(l\wedge p)\sin(-l\wedge p - l\wedge q)\sin(l\wedge q + p\wedge q), \nonumber
\end{eqnarray}

where their PP contributions are
\begin{eqnarray}
&& {\rm PP}[B^\lambda_{\;\;\nu'}(p,q,k)(q^2 g^{\nu\nu'} - q^\nu q^{\nu'}) + B^\nu_{\;\;\lambda'}(p,k,q)(k^2 g^{\lambda\lambda'} - k^\lambda k^{\lambda'})] \nonumber\\ 
&&\quad\quad\quad\quad\quad\quad = -\frac{e^2}{16\pi^2}\frac{1}{\epsilon}[(k^2 g^{\lambda\nu} - k^\lambda k^\nu) - (q^2 g^{\nu\lambda} - q^\nu q^\lambda)].
\end{eqnarray}
From this, we see that the renormalization constant $\tilde Z_1$ must be
\begin{equation}
\tilde Z_1= 1-\frac{e^2}{16\pi^2} \frac{1}{\epsilon},
\end{equation}
so that $\tilde Z_1=Z_5$ and thus we obtain the relations
\begin{equation}
Z_2/Z_{1F} = \tilde Z_3/\tilde Z_1 = Z_3/Z_1.
\end{equation}

Therefore, the sum of all PP contributions of the right-hand side of Eq.~(\ref{ITF}) becomes
\begin{eqnarray}
&& PP[\Pi^{\nu\lambda}_{b,c,d}(q) - \Pi^{\lambda\nu}_{b,c,d}(k)+{B^{\lambda}}_{\nu'}(p,q,k)(q^2 g^{\nu\nu'} - q^\nu q^{\nu'}) + {B^{\nu}}_{\lambda'}(p,q,k)(k^2 g^{\lambda\lambda'} - q^\lambda q^{\lambda'})] \nonumber\\ 
&&\quad\quad\quad\quad\quad\quad = \frac{7e^2}{48\pi^2}\frac{1}{\epsilon}[(k^2 g^{\lambda\nu} - k^\lambda k^\nu) - (q^2 g^{\nu\lambda} - q^\nu q^\lambda)],
\end{eqnarray}
which is the same result as for the left-hand side, Eq.~(\ref{sumL}).

Besides these ultraviolet divergent parts, arising from the planar parts of the diagrams, we have also infrared singular parts (SP) coming from the nonplanar parts of the same diagrams, at $p,q,k=0$. Explicit calculations, combining denominators with Feynman parameters and using nonplanar integrals, give us the SP for the diagrams $\Lambda^{\mu\nu\lambda}$ on the left-hand side of Eq.~(\ref{ITF}): 
\begin{equation}
{\rm SP}[2e\sin(p\wedge q)\Lambda_{b1,b2,c,d1,d2,d3}^{\mu\nu\lambda}(p,q,k)] = \frac{4e^3}{\pi^2}\frac{\sin(p\wedge q)}{p\wedge q}\left(\frac{\tilde p^\mu\tilde p^\nu\tilde p^\lambda}{\xi\,\tilde p^4}+\frac{\tilde q^\mu\tilde q^\nu\tilde q^\lambda}{\xi\,\tilde q^4}+\frac{\tilde k^\mu\tilde k^\nu\tilde k^\lambda}{\xi\,\tilde k^4}\right),
\end{equation}
where we are not taking into account the logarithmic singularities. Contracting $q_\mu$ in the above expression, we obtain
\begin{equation}
{\rm SP}[q_\mu\Lambda_{b1,b2,c,d1,d2,d3}^{\mu\nu\lambda}(p,q,k)] = \frac{2e^2}{\pi^2}\left(\frac{\tilde q^\nu \tilde q^\lambda}{\xi^2\tilde q^4}-\frac{\tilde k^\nu \tilde k^\lambda}{\xi^2\tilde k^4}\right),
\end{equation}
which is exactly the same SP for the photon self-energy diagrams on the right-hand side of Eq.~(\ref{ITF}),
\begin{eqnarray}
&& {\rm SP}[\Pi^{\nu\lambda}_{b,c,d}(q) - \Pi^{\lambda\nu}_{b,c,d}(k)] = \frac{2e^2}{\pi^2}\left(\frac{\tilde q^\nu \tilde q^\lambda}{\xi^2\tilde q^4}-\frac{\tilde k^\nu \tilde k^\lambda}{\xi^2\tilde k^4}\right).
\end{eqnarray}
The other diagrams of the ST identity (\ref{ITF}) contribute only with logarithmic SP. These singularities are not problematic as they are integrable. 

\section{Final comments}

In this work, for some specific Green functions,  we have analysed the ST identities in the context of noncommutative QED$_4$. Special attention was given to the vectorial fermion-photon and triple photon vertex functions, explicitly verifying that no anomalies arise. The validity of these identities imply that, in spite of the presence of dangerous infrared singularities, the ultraviolet structure is not essentially modified by the noncommutativity.  In fact, although the individual pole parts have been changed and new divergences appeared, the counterterms are related as they should in a non-Abelian situation. This however does not preclude the occurrence of dangerous infrared singularities which, in higher orders,  jeopardizes the perturbative series. To extend our results to higher orders, our study must therefore be supplemented by some mechanism to control the mentioned singularities. 
One possibility is to consider the effect of supersymmetry; as known supersymmetric theories have a better ultraviolet behavior and consequently
they may be free from dangerous infrared/ultraviolet mixing. This is what happens in susy noncommutative QED$_4$ \cite{ferrari}.

\end{document}